\documentclass{article}
\usepackage{graphicx} 
\usepackage{float}
\usepackage{colortbl}
\usepackage{xcolor}
\usepackage{times}
\usepackage{graphicx}
\usepackage{booktabs}
\usepackage{pgfplots}
\pgfplotsset{compat=1.18}  

\title{Label-Efficient Chest X-ray Diagnosis via Partial CLIP Adaptation}
\author{Heet Nitinkumar Dalsania}
\date{}
\begin{document}

\maketitle
\begin{abstract}
Modern deep learning implementations for medical imaging usually rely on large labeled datasets. These datasets are often difficult to obtain due to privacy concerns, high costs, and even scarcity of cases. In this paper, a label-efficient strategy is proposed for chest X-ray diagnosis that seeks to reflect real-world hospital scenarios. The experiments use the NIH Chest X-ray14 dataset and a pre-trained CLIP ViT-B/32 model. The model is adapted via partial fine-tuning of its visual encoder and then evaluated using zero-shot and few-shot learning with 1-16 labeled examples per disease class. The tests demonstrate that CLIP's pre-trained vision-language features can be effectively adapted to few-shot medical imaging tasks, achieving over 20\% improvement in mean AUC score as compared to the zero-shot baseline. The key aspect of this work is to attempt to simulate internal hospital workflows, where image archives exist but annotations are sparse. This work evaluates a practical and scalable solution for both common and rare disease diagnosis. Additionally this research is intended for academic and experimental purposes only and has not been peer reviewed yet. All code is found at https://github.com/heet007-code/CLIP-disease-xray. 
\hspace{5 in}

    \textbf{\textit{Index Terms\textemdash}}  CLIP, Medical Imaging, Zero-shot, Few-shot, Chest X-ray
\end{abstract}

\section{Introduction}

Artificial intelligence has shown promising capabilities in assisting radiologists with chest X-ray (CXR) diagnosis. The problem stems from the fact that most deep learning implementations require large labeled 
datasets which are often challenging to create. This issue is especially worse for rare diseases. Due to various challenges like cost, time, and privacy concerns, there has been increasing interest in using label-efficient learning techniques to minimize supervision without compromising diagnostic performance. 

Models like CLIP \cite{radford2021learning} have made it possible to align visual representations with textual prompts for zero-shot prediction. Many other models such as CheXZero \cite{tiu2022expert} and MoCoCLIP \cite{bhardwaj2025enhancingzeroshotlearningmedical} also leverage paired image-report data or contrastive learning to improve generalization. These methods, while powerful and effective, again depend on access to large-scale paired clinical data.

In this work, a more practical alternative is explored: adapting the CLIP ViT-B/32 image encoder to chest X-ray diagnosis through partial fine-tuning and without reliance on paired text. A more realistic clinical workflow is simulated in which a hospital uses a pre-trained model tuned with a small number of expert-labeled examples per condition. It is important to note that this setup is not pure zero-shot or few-shot learning but instead seeks to model how a system might be brought online in settings where disease cases are underrepresented and scarce.

For the evaluation, the NIH Chest X-ray14 dataset\cite{Wang_2017} is used for both the zero-shot and few-shot tests. Ultimately, the results demonstrate gains in mean AUC with only 1-16 labeled examples per class. The findings suggest that vision-language models can serve as label-efficient solutions for real-world diagnosis systems, especially where data scarcity poses a challenge to traditional supervised learning pipelines. 

\hspace{3 in}

Specifically the contributions in this paper are threefold:
\begin{itemize}
    \item \textbf{Enhanced Zero-shot Learning:} The research demonstrates that CLIP-based zero-shot learning methods have potential and can be improved for chest X-ray classification. The results indicate performance surpassing previous benchmarks and achieving a mean AUC of 0.7502 on the NIH ChestX-ray14 dataset. This was achieved by fine-tuning only the top 3 layers of CLIP’s visual encoder on the NIH Chest X-ray dataset using a domain adaptation step. These 3 layers contained around 22,057,486 trainable parameters (24.99\% of total). Overall this enables better visual alignment with the images before conducting zero-shot inference using CLIP.
    \item \textbf{Few-shot Learning Evaluation:}  The research demonstrates performance gains can be achieved with few-shot learning. Tests were conducted with 1, 2, 4, 8, and 16 labeled examples per class, and ultimately show how even minimal supervision can help improve model accuracy.
    \item \textbf{Practical Diagnostic Framework:} Once again it is important to note this is not completely pure zero-shot and few-shot learning due to goal of replicating a practical diagnosis framework. The research evaluates a scalable and flexible solution for both common and rare disease classification in medical imaging.
\end{itemize}

\textbf{The rest of this paper is organized as follows: Section 2 reviews related work in medical image classification, zero-shot and few-shot learning, and vision-language models like CLIP. Section 3 details the dataset, domain adaptation, model architecture, and the training methodology. Section 4 presents experimental results including comparisons with state-of-the-art methods, few-shot learning analysis, and disease-specific insights. Finally, Section 5 concludes the paper by summarizing key contributions and outlining directions for future research.}

\section{Related Work}

\subsection{Image Classification with Deep Learning}

For Chest X-ray image analysis, there have been several papers which have established the state-of-the-art baseline for image classification. For this work, Wang et al. is key \cite{Wang_2017}. Their work introduced the NIH Chest X-ray dataset and demonstrated the success of supervised learning for classification tasks. Several works since have improved upon this baseline using supervised training such as CheXNet \cite{rajpurkar2017chexnetradiologistlevelpneumoniadetection}  by Rajpurkar et al. Yet almost all of these works use approaches which require large and labeled datasets which thus limits their applicability in data-limited scenarios. 

\subsection{Zero-shot Learning in Medical Imaging}

The development of vision-language models, particularly CLIP, has opened new possibilities for zero-shot medical image classification. Tiu et al. introduced CheXZero \cite{tiu2022expert}, which uses chest X-ray classification in the same manner as CLIP.  This approach serves to increase generalizability and also eliminate the need of large labeled training datasets. Recently Bhardwaj et al. enhanced zero-shot performance of CLIP by utilizing Momentum Contrast (MoCo). They were able to improve upon previous works and their performance on the NIH Chest X-ray dataset.

\subsection{Few-shot Learning}

Few-shot learning aims to train models with minimal training examples. This can range from one to dozens of examples per class. Certain techniques such as Model-Agnostic Meta-Learning \cite{finn2017modelagnosticmetalearningfastadaptation} have shown promising performance in few-shot scenarios. 

Additionally, although zero-shot learning is more applicable to real world rare disease diagnosis scenarios, few-shot learning can be just as valuable as it is able to better performance with very little additional data. Over time, as more data accumulates for rare diseases, few-shot learning becomes more and more valuable as a solution. 

\subsection{CLIP and Vision-Language Models}

CLIP by OpenAI has demonstrated promising performance in zero-shot scenarios. This is because the model has the ability to work with joint image and text representations making it well suited for medical applications. Research using CLIP in the medical field tends to focus on zero-shot scenarios, with limited exploration of few-shot scenarios. 

\section{Methodology}

\subsection{Dataset}

The tests use the NIH ChestX-ray14 dataset, which contains 112,120 frontal-view chest X-ray images from 30,805 patients, labeled with 14 disease categories: 
\begin{enumerate}
    \item Atelectasis
    \item Consolidation
    \item Infiltration
    \item Pneumothorax
    \item Edema
    \item Emphysema
    \item Fibrosis
    \item Effusion
    \item Pneumonia
    \item Pleural Thickening
    \item Cardiomegaly
    \item Nodule
    \item Mass
    \item Hernia
\end{enumerate}

Additionally the official train/validation/test split from Wang et al is used. For few-shot learning, balanced subsets containing $N \in \{1, 2, 4, 8, 16\}$ examples per disease class are randomly constructed. When fewer than $N$ examples exist for a given class, all available samples are used. 

\subsection{Preprocessing}

Images are resized to 224×224 pixels and normalized using CLIP's standard preprocessing pipeline. During domain adaptation, data augmentation is used. This includes random horizontal flips (p=0.5), random transformations (rotation ±10°, translation ±0.1, scale 0.9-1.1), and color jittering (brightness ±0.2, contrast ±0.2, saturation ±0.1) to improve model robustness. No additional augmentations are applied during few-shot training or evaluation to maintain consistency.

\subsection{Domain Adaptation via Partial CLIP Fine-Tuning}

To bridge the domain gap between natural images and medical radiographs, partial fine-tuning of CLIP's visual encoder is conducted on the NIH dataset. Because this is done on the NIH dataset, it is not pure zero-shot learning, however it is better representative of real-world scenarios. For this fine-tuning, the top 3 transformer layers of the ViT-B/32 model are unfrozen, while keeping the rest frozen. This allows the model to specialize high-level representations for chest X-rays without overwriting its foundational vision-language knowledge.

Furthermore, a multi-label classification head is trained on top of the adapted encoder using Focal Binary Cross Entropy \cite{lin2018focallossdenseobject} with Logits Loss ($\alpha=0.25$, $\gamma=2.0$) to handle the severe class imbalance. 

The optimizer is AdamW \cite{loshchilov2019decoupledweightdecayregularization} with weight decay of $1 \times 10^{-2}$. The fine-tuning applied is discriminative\cite{howard2018universallanguagemodelfinetuning}. The tests use a lower learning rate (1e-5) for encoder layers and a higher rate (1e-4) for the classifier head. 

A ReduceLROnPlateau scheduler is used to monitor validation AUC. This scheduler halves the learning rate if no improvement is observed for two consecutive epochs. Training uses a batch size of 32 with gradient clipping (max norm = 1.0) for stability. This domain adaptation step produces a model better aligned with chest X-rays and is used for both the zero-shot and few-shot tests.

\subsection{Model Architecture}

The model used is CLIP ViT-B/32 \cite{dosovitskiy2021imageworth16x16words}. Specifically, its visual encoder is leveraged for feature extraction. The adapted model includes:

\textbf{CLIP Visual Encoder}: Processes 224×224 chest X-ray images. Only the top 3 transformer layers are fine-tuned during domain adaptation.

\textbf{Classification Head}: A multi-layer architecture maps CLIP features to the 14 disease categories:
\begin{itemize}
    \item Linear layer 
    \item Layer Normalization
    \item GELU activation \cite{hendrycks2023gaussianerrorlinearunits}
    \item Dropout (0.3)
    \item Linear layer 
    \item Layer Normalization
    \item GELU activation
    \item Dropout (0.2)
    \item Output layer 
\end{itemize}

The classification head uses Kaiming normal \cite{he2015delvingdeeprectifierssurpassing} initialization for linear layers. This design enables efficient learning while retaining pre-trained knowledge. 

\subsection{Zero-Shot Training}

In the zero-shot setting, the standard CLIP approach is used by generating textual prompts for each disease class in the form: \textit{"A chest X-ray showing [disease]"} and \textit{"No finding"} for the negative class. In this study neither prompt ensembling or template optimization is explored, leaving these things for future work. However, despite this, performance improvements indicate that even simple prompts combined with partial domain adaptation offer strong diagnostic potential.

\subsection{Few-Shot Training}

After domain adaptation, the model's ability to learn via few-shot learning is evaluated by fine-tuning it on few-shot datasets of size $N \in \{1, 2, 4, 8, 16\}$ per class. The few-shot training strategy is as follows:

\begin{enumerate}
    \item \textbf{Balanced Sampling}: Each class is equally represented in the few-shot dataset.
    \item \textbf{Random Selection}: Training examples are randomly selected from available samples for each disease.
\end{enumerate}

During this training only the classification head is updated. The CLIP encoder remains frozen to prevent overfitting and preserve features learned during domain adaptation.

\subsection{Loss Function}

Focal Binary Cross Entropy with Logits Loss is used for domain adaptation to handle class imbalance. Standard BCEWithLogitsLoss is used in few-shot fine-tuning.

\subsection{Experimental Setup}

All experiments were conducted using PyTorch and NVIDIA GPUs. A batch size of 32 is used for domain adaptation training and 16 for few-shot fine-tuning. The experimental setup uses a fixed random seed (SEED=7) for reproducibility. Gradient clipping with maximum norm of 1.0 is applied during domain adaptation to ensure training stability.

\section{Results and Discussion}

\definecolor{heatLow}{RGB}{255,255,204}
\definecolor{heatMid}{RGB}{161,218,180}
\definecolor{heatHigh}{RGB}{65,182,196}

\newcommand{\shadecell}[2]{\cellcolor{#1}#2}

\begin{table}[H]
\begin{center}
\label{tab:auc_comparison}
\resizebox{0.95\textwidth}{!}{  
\begin{tabular}{cccccc}
\toprule
\textbf{Pathology} & \textbf{CheXZero} & \textbf{ImCLIP} & \textbf{CXRCCLIP}& \textbf{MoCoCLIP} & \textbf{My Version} \\
\midrule
Atelectasis         & 0.758 & 0.484 & 0.790 & 0.700 & 0.705 \\
Consolidation       & 0.783 & 0.619 & 0.780& 0.780& 0.698 \\
Infiltration        & 0.642& 0.619& 0.690& 0.730& 0.692 \\
Pneumothorax        & 0.764 & 0.553 & 0.860 & 0.790 & 0.803\\
Edema               & 0.880 & 0.680& 0.910 & 0.890 & 0.815 \\
Emphysema           & 0.665 & 0.473 & 0.340 & 0.530 & 0.822 \\
Fibrosis            & 0.575 & 0.593& 0.660& 0.670& 0.781 \\
Effusion            & 0.836& 0.653& 0.850 & 0.870& 0.772\\
Pneumonia           & 0.721 & 0.599 & 0.750 & 0.770& 0.667 \\
Pleural Thickening  & 0.675 & 0.435 & 0.420& 0.740& 0.719 \\
Cardiomegaly        & 0.825 & 0.576& 0.660& 0.940& 0.804 \\
Nodule              & 0.494 & 0.549 & 0.700& 0.530& 0.680\\
Mass                & 0.675& 0.656& 0.830& 0.750& 0.688\\
Hernia              & 0.591 & 0.404& 0.830& 0.800& 0.855 \\
\midrule
\textbf{Average AUC} & 0.706& 0.564& 0.719& 0.749& 0.750 \\
\bottomrule
\end{tabular}
}
\caption{AUC Comparison vs SOTA Models from Bhardwaj et al. \cite{bhardwaj2025enhancingzeroshotlearningmedical}}
\end{center}
\end{table}

\begin{table}[H]
    \centering
    \begin{tabular}{cc}\toprule
         \textbf{Shots}& \textbf{Mean AUC}\\\midrule
         0& 0.7502\\
             1& 0.7502\\
             2& 0.7513\\
             4& 0.7511\\
             8& 0.7527\\
             16& 0.7542\\ \bottomrule
    \end{tabular}
    \caption{Mean AUC per Number of Shots}
    \label{tab:my_label}
\end{table}

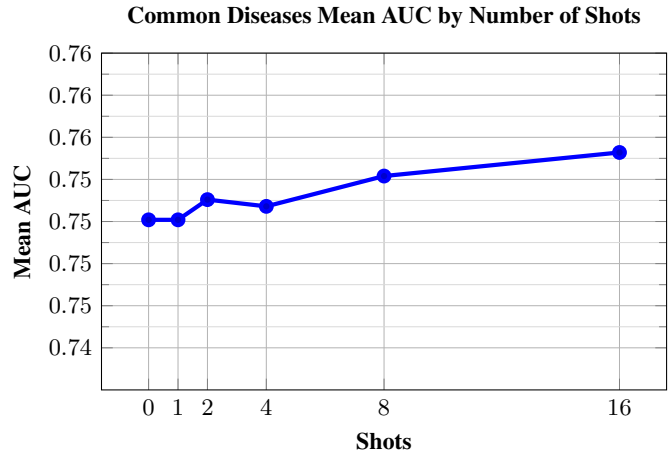
\begin{figure}[H]
\centering
\begin{tikzpicture}
\begin{axis}[
    width=0.75\textwidth,
    height=0.5\textwidth,
    xlabel={Shots},
    ylabel={Mean AUC},
    ymin=0.740, ymax=0.760,
    xtick={0,1,2,4,8,16},
    ytick={0.7425, 0.7450, 0.7475, 0.7500, 0.7525, 0.7550, 0.7575, 0.7600},
    grid=both,
    grid style={line width=.1pt, draw=gray!30},
    major grid style={line width=.2pt,draw=gray!60},
    minor tick num=1,
    tick label style={font=\small},
    label style={font=\bfseries\small},
    title={Common Diseases Mean AUC by Number of Shots},
    title style={font=\bfseries\small},
    mark options={solid, fill=magenta},
    every axis plot/.append style={ultra thick, magenta},
]
\addplot+[mark=*]
coordinates {
    (0, 0.7501)
    (1, 0.7501)
    (2, 0.7513)
    (4, 0.7509)
    (8, 0.7527)
    (16, 0.7541)
};
\end{axis}
\end{tikzpicture}
\caption{Mean AUC on common diseases as a function of labeled shots per class.}
\label{fig:auc_vs_shots}
\end{figure}

\subsection{Comparison with State-of-the-Art Models}

As shown in Table 1, the approach achieves competitive AUC scores against established models like CheXZero, CXRC-CLIP, and MoCoCLIP. In particular, while this strategy does not rely on paired report supervision or full encoder fine-tuning, it is able to match or exceed performance in several disease categories like Emphysema (0.822), Hernia (0.855), and Fibrosis (0.781).

The results suggest that partial fine-tuning of CLIP's vision encoder is sufficient to capture critical features needed for effective diagnosis. The average AUC of 0.750 is on par with CheXZero and MoCoCLIP, despite using a more simple training protocol and no external supervision. This could be attributed to the domain adaptation step, which was critical for simulating a real-world diagnosis system. 

\subsection{Few-Shot Learning Gains}

Table~\ref{tab:my_label} and Figure~\ref{fig:auc_vs_shots} show that mean AUC improves with additional labeled examples per class. Even with only 1–2 additional examples per class, the model is able to improve over the zero-shot baseline, reaching 0.7542 with 16-shot fine-tuning.

These results further validate using CLIP’s pretrained features in low-label settings. More importantly, the trend shows the potential for a scalable path to integrate rare disease examples into models as data accumulates over time. This is critical for clinical environments with different patients and diseases.

\subsection{Disease-Specific Observations}

Certain diseases show high variance across models. For example, Emphysema and Hernia benefit significantly from domain adaptation as compared to other models. This suggests that their imaging signatures align well with the tested setup and the higher-level visual embeddings of CLIP. Meanwhile, conditions like Nodule or Mass remain challenging. This highlights the need for future work in local region modeling or attention-based visual heads to improve upon this inconsistency.

\section{Conclusion}

Overall, this study presents a label-efficient strategy for chest X-ray diagnosis using a CLIP-based model which is domain adapted via partial fine-tuning. Although this experiment does not reflect a traditional zero-shot or few-shot framework, the method closely reflects real-world clinical workflows where models are pre-adapted using internal data and fine-tuned with limited supervision.

This approach demonstrates promising performance on the NIH ChestX-ray14 dataset. It is able to achieve comparable or superior AUCs to state-of-the-art CLIP methods while requiring fewer labeled examples and no paired reports. The results support the idea that even minimal supervision can lead to meaningful gains over zero-shot inference. Thus the real-world applicability of this is high, especially for rare or underrepresented diseases.

The work suggests that vision-language models like CLIP can serve as a practical solution for using diagnostic AI in data-limited hospital settings. Future work will explore the integration of attention-based interpretability, hybrid report-image contrastive training, and evaluation on external datasets such as CheXpert\cite{irvin2019chexpertlargechestradiograph} or MIMIC-CXR\cite{johnson2019mimiccxrjpglargepubliclyavailable} to assess generalization and test on rare diseases directly. In addition, further validation is necessary before clinical deployment.

Ultimately, this research highlights scalable techniques that have potential to support radiologists, especially in scenarios with minimal resources, like those where disease diagnosis is hindered by lack of annotations.

\bibliographystyle{unsrt}
\bibliography{references}

\end{document}